\documentclass[12pt,preprint]{aastex}

\begin{document}

%\documentclass[12pt,preprint]{aastex}
%\usepackage{natbib}
%\citestyle{aa}

%\received{}
%\accepted{}
%\journalid{}{}
%\articleid{}{}
\title{On the properties of thermal disk winds in X-ray transient sources:
a case study of GRO J1655--40.}

\author{
S. Luketic\altaffilmark{1}, 
D. Proga\altaffilmark{1}, 
T.R. Kallman\altaffilmark{2}, 
J.C. Raymond\altaffilmark{3}, and
J.M. Miller\altaffilmark{4}}

\affil{$^{1}$ Department of Physics, University of Nevada, Las Vegas,
NV 89154} \email{\{stefan, dproga\}@physics.unlv.edu}
\affil{$^2$ NASA, Goddard Space Flight Center, Laboratory for High Energy Astrophysics, Code 662, Greenbelt, MD 20771}
\affil{$^3$ Harvard-Smithsonian Center for Astrophysics, 60 Garden St., 
Cambridge, MA 02138}
\affil{$^4$ Department of Astronomy, University of Michigan, 500 Church Street, Ann Arbor, MI 48109}

\def\LSUN{\rm L_{\odot}}
\def\MSUN{\rm M_{\odot}}
\def\RSUN{\rm R_{\odot}} 
\def\MSUNYR{\rm M_{\odot}\,yr^{-1}}
\def\MSUNS{\rm M_{\odot}\,s^{-1}}
\def\MDOT{\dot{M}}

\newbox\grsign \setbox\grsign=\hbox{$>$} \newdimen\grdimen \grdimen=\ht\grsign
\newbox\simlessbox \newbox\simgreatbox
\setbox\simgreatbox=\hbox{\raise.5ex\hbox{$>$}\llap
     {\lower.5ex\hbox{$\sim$}}}\ht1=\grdimen\dp1=0pt
\setbox\simlessbox=\hbox{\raise.5ex\hbox{$<$}\llap
     {\lower.5ex\hbox{$\sim$}}}\ht2=\grdimen\dp2=0pt
\def\simgreat{\mathrel{\copy\simgreatbox}}
\def\simless{\mathrel{\copy\simlessbox}}

\begin{abstract}
We present the results of hydrodynamical simulations of the disk photosphere 
irradiated by strong X-rays produced in the inner-most part of the disk of an accreting black hole.
As expected, the irradiation heats the photosphere and drives a thermal wind.
To apply our results to the well-studied X-ray transient source GRO J1655--40, 
we adopted the observed mass of its black hole, and the observed properties 
of its X-ray radiation. To compare the results with the observations, we also 
computed transmitted X-ray spectra based on the wind solution. 
Our main finding is: 
the density of the fast-moving part of the wind is more than one order 
of magnitude lower than that inferred from the observations. 
Consequently, the model fails to predict spectra with line absorption as 
strong and as blueshifted as those observed. However, despite the thermal 
wind being weak and Compton thin, the ratio between the mass-loss rate and 
the mass accretion rate is about seven. This high ratio is insensitive 
to the accretion luminosity, in the limit of lower luminosities. Most of 
the mass is lost from the disk between 0.07 and 0.2 of the Compton radius. 
We discovered that beyond this range  the wind solution is self-similar. 
In particular, soon after it leaves the disk, the wind flows at a constant 
angle with respect to the disk. Overall, the thermal winds generated in our 
comprehensive simulations do not match the wind spectra observed in 
GRO J1655--40.  This supports the conclusion of Miller et al. and 
Kallman et al. that the wind in GRO J1655--40, and possibly other X-ray 
transients, may be driven by magnetic processes. This in turn implies that 
the disk wind carries even more material than our simulations predict and 
as such has a very significant impact on the accretion disk structure 
and dynamics. 

\end{abstract}

\keywords{accretion, accretion disks   --
methods: numerical -- HD}

\section{INTRODUCTION} 

Most X-ray sources are powered by disk accretion 
onto compact objects.
Therefore a main challenge
for X-ray astronomy is to understand 
the mechanisms that enable this process. High quality and 
high spectral-resolution
observations obtained with {\it Chandra}, {\it XMM-Newton}, and {\it Suzaku} 
allow us to 
study disk accretion and related outflows
better than ever before. These observations are especially
revealing if taken of relatively bright, well studied objects such as
X-ray transient sources, with GRO~J1655--40 being
a very fine example. 

Many properties of GRO~J1655--40 are well constrained. For example,
GRO J1655--40 is a black hole binary at a distance of 3.2 kpc 
containing a black hole with a mass of 7.0 $\MSUN$ that accretes from 
an F3 IV--F6 IV star with a mass of 2.3 $\MSUN$ in a 2.6-day orbit. 
The inner disk is viewed at an inclination of $67^\circ$ -- $85^\circ $ (nearly edge-on; Orosz $\&$ Bailyn 1997). 
When GRO~J1655--40 is in an X-ray--bright phase, it is 
possible to obtain with {\it Chandra} high signal-to-noise spectra 
as shown by Miller et al. (2006a; M06 hereafter). In fact, the quality of
the spectra obtained by M06 is good enough to reveal many absorption
lines significant at the 5 $\sigma$ level of confidence or higher. 
Over 70 of these lines can be identified as resonance lines from 
over 32 charge states. 
This is in
sharp contrast to high resolution spectra of other
systems (see below) and of GRO J1655--40 in other spectral states,
when only the absorption lines of Fe XXV and XXVI
are detected (e.g., Miller et al. 2006b; Kubota et al. 2007;
Miller et al. 2008; Nielsen \& Lee 2009).

It is hoped that such high quality spectroscopy can provide new 
and surprising results. Indeed, absorption lines discovered by M06
are blueshifted and likely produced in a disk wind, making GRO~J1655--40  
the best case for an X-ray binary with an X-ray absorbing disk wind.
But there are also other cases.
For example, {\it Chandra} and {\it RTXE} spectroscopy
of the microquasar H~1743$-$322 reveals blueshifted absorption lines
that are likely formed in a disk wind
(Miller et al. 2006b). Another example is the X-ray transient 4U 1630-472.
Outburst spectra of this source obtained with {\it Susaku}, show
iron absorption lines indicative of a disk wind (Kubota et al. 2007).
More recently, Neilsen \& Lee (2009) found an X-ray absorbing disk wind
in the microquasar GRS 1915+105.
In addition, there is an outflowing X-ray absorber in  Circinus X-1,
which could either be a disk wind  or  a wind 
from a massive companion star (Schulz \& Brandt 2002).

The disk wind discovered in GRO~J1655--40 gives us a good 
testbed for constraining wind properties using X-ray observations.
For example, by fitting to observations 1-dimensional (1-D) models, M06
and Miller et al. (2008)
put strict limits on the ionization balance in the wind. 
This was facilitated by the fact
that some of the detected lines provide a density diagnostic.
The main results from the fitting yielded the following parameters:
number density $n$ between $5\times10^{13}~\rm{cm^{-3}}$ and 
$2\times10^{14}~\rm{cm^{-3}}$, 
column density $N_{\rm H}\,=\,7.4\times 10^{21}~\rm{cm}^{-2}$, 
distance from the central object to the point of emerging of the strong wind 
$4.8\times10^8$~cm and velocity $v_{\rm r}\,=\,500~\rm{km\,s}^{-1}$. 
The lines show blueshifts in the 300--1600~$\rm{km~s^{-1}}$ range. 

Using these wind properties, together with the observed wind speed and
the system luminosity, M06 concluded that 
the inferred wind location is well within the Compton radius 
(the radius where the gravitational and thermal pressures are equal), 
defined by: 
\begin{equation}
R_{\rm IC} = \frac{G M_{\rm BH} m_{\rm p} \mu}{k T_{{\rm IC}} } ,
%= \frac{1.0 \times 10^{10}}{T_{C8}} \frac{M}{M_{\odot}} \,\, {\rm cm} \,\, ,
\end{equation}
where $M_{{\rm BH}}$ is the mass of the black hole, 
$\mu$ is the mean molecular weight, $m_{\rm p}$ is the proton mass, 
and $T_{{\rm IC}}$ is the Compton temperature (for $M_{{\rm BH}}=7\, \MSUN$, 
$\mu\,=\,0.6$, and $T_{{\rm IC}}=1.4\times10^7$~K, the Compton
radius is $R_{\rm IC}=4.8\times 10^{11}$ cm).
This wind location is then inconsistent with an outflow being 
driven by thermal expansion, 
even though weak flows are possible at $\sim$ 0.1 of the Compton radius, 
as found by Woods et al. (1996; see also Proga \& Kallman 2002, hereafter PK02).
On the other hand, the wind cannot be driven by
radiation pressure in this system because the luminosity, $L_\ast$, relative to
the Eddington limit, $L/L_{\rm Edd}=0.03$, is too low (e.g., PK02). 
Therefore, by a process of elimination, M06 concluded
that the only plausible mechanism that could drive the wind is
magnetic processes. MO6 also concluded that disk accretion itself is driven 
by magnetic fields. 

Although magnetic forces can drive a disk wind
(e.g., Blandford \& Payne 1982) and magnetic fields
are a very important ingredient of accretion disks 
(Balbus \& Hawley 1998), M06's arguments for magnetic driving
are indirect (see also Proga 2006).  Ideally, one should demonstrate
that magnetic forces can drive a disk wind capable of reproducing
the observed spectra. However, that is a relatively challenging task, 
and instead one could try to verify first whether thermal driving is indeed 
unsatisfactory.
This is a very relevant question especially in light of the work done by
Netzer (2006; N06 hereafter), who argued that it is possible to  produce 
a simple thermally driven wind that is consistent with the observed 
properties of GRO~J1655--40.

One can argue for or against thermal driving by comparing
the escape velocity and the isothermal
sound speed at the Compton temperature 
(equivalently, one can compare the wind launching radius, $R_{\rm l}$ with, 
$R_{{\rm IC}}$ as M06 did).
To use this basic physical argument, one needs to estimate 
the wind temperature and $R_{\rm l}$. 
The former can be constrained relatively well for the Compton heated gas, 
whereas the latter is much more difficult to constrain.
In fact, the controversy about the role of thermal driving has its origin
in two different estimates for $R_{\rm l}$: 
M06 estimated $R_l=5\times10^{8}$ cm, whereas N06 estimated $R_l=5\times10^{10}-5\times10^{11}$ cm. 
These two estimates differ  by
2--3 orders of magnitude! For the lower estimate, the wind
cannot be thermally driven, whereas for the higher estimate it can.
We note that the two groups also found different values
for $R_{\rm IC}$ (i.e., M06's estimate is $7\times10^{12}$~cm
whereas N06's estimate is $5\times10^{11}$~cm). 
Additionally, the two groups disagree about the gross
properties of the thermal wind. In particular, contrary to N06, M06
claimed that both theory and simulations
(Begelman et al. 1982; Woods et al. 1996) predict mass-loss rates 
of thermal winds that are much too low to account for the observed density
and blueshift.

The reason for the above disagreement can be traced down to the differences 
about the absorption lines of $\rm{Fe\,XXII}\,and \,\rm{Fe\,XXIII}$. 
N06 assumed saturated lines and 
the absolute covering factor 
of the X-ray source to be 0.75, whereas M06 took the lines to be unsaturated
and used a covering factor of 1.

Miller et al. (2008) confirmed their previous results and  stated that if
NO6's claim of the $\rm{Fe\,XXII}$ line being saturated were correct then
the ratios of some other lines should be different than what is observed. 
Miller et al. also re-iterated their point that 
that $\rm{Fe\,XXII}$ lines are optically thin and that 
the covering factor along the line of the sight is near unity. 
They also stated that N06 used too low a density 
(by factor of 5--10) as a consequence of  
his line saturation assumption.

The conclusion of M06 and Miller et al. (2008) was confirmed by
Kallman et al. (2009) who used spectral fitting to show
that the ionization conditions in a 1-D model are not consistent 
with the wind being driven by thermal expansion. 

Generally, M06, N06, and Kallman et al. (2009) agree that thermal driving 
operates in this system but disagree about the properties of the thermal wind. 
Overall, it appears that thermal driving is unfavorable. However, 
this conclusion is based on simplified 1-D wind models and needs to be 
confirmed by a detailed physical model that takes into account 
the intrinsic multi-dimensional geometry of disk winds.

In this paper, we present 2.5-D axisymmetric, time-dependent hydrodynamical 
numerical simulations of thermally driven disk winds. We focus on the results
obtained after the simulations reached the steady state and compare
the results with observations. Our simulations do not include any magnetic 
processes and therefore do not address the possibility of the wind being 
magnetically driven, but they instead focus on the thermal contribution to 
the wind. The main goal of our simulations is to assess within 
the assumptions whether or not the thermally driven wind can account for 
the observations. 
We also consider possible implications of the fact that
the mass-loss rate of the wind is several times higher than
the total mass-accretion rate.

The paper is organized in the following way: Section 2 describes 
the methods used; Section 3 lays out the initial conditions of 
the simulation and discusses the properties of the fiducial run as well 
as its differences from 
other runs; Section 4 compares the results to the observations; 
Section 5 summarizes and discusses our results.

\section{METHOD}

\subsection{HYDRODYNAMICS}

To compute the structure and evolution of a disk irradiated by the central 
source, we numerically solve the equations of hydrodynamics 
\begin{equation}
   \frac{D\rho}{Dt} + \rho \nabla \cdot {\bf v} = 0,
\end{equation}
\begin{equation}
   \rho \frac{D{\bf v}}{Dt} = - \nabla P + \rho {\bf g}
% + \rho {\bf F}^{\rm rad}
\end{equation}
\begin{equation}
   \rho \frac{D}{Dt}\left(\frac{e}{ \rho}\right) = -P \nabla \cdot {\bf v}
   + \rho \cal{L}  ,
\end{equation}
where $\rho$ is the mass density, $P$ is the gas pressure,
${\bf v}$ is the velocity, $e$ is the internal energy density,
$\cal{L}$ is the net cooling rate, and
${\bf g}$ is the gravitational acceleration of the central object.
% ${\bf F}^{\rm rad}$ is the total radiation force per unit mass.
We adopt an adiabatic equation of state, $P~=~(\gamma-1)\,e$, 
and consider models with the adiabatic index, $\gamma=5/3$.

We use the ZEUS-2D code (Stone \& Norman 1992)
extended by Proga et al. (2000) to solve eqs. 2--4. 
We perform our calculations in spherical polar coordinates
$(r,\theta,\phi)$ assuming axial symmetry about the rotational axis
of the accretion disk ($\theta=0^o$). 

Our computational domain is defined to occupy 
the angular range $0^o \leq \theta \leq 90^o$ and the radial range
$r_{\rm i}\leq r \leq \ r_{\rm o}$. The
numerical resolution consists of 200 cells in the $r$ direction and
100 cells in the $\theta$ direction. 
In the angular direction, we used the following ratios: 
$d{\theta}_{{\rm k+1}}/d{\theta}_{{\rm k}}$=0.95, 0.97, 0.99, and 1.00
[i.e., the zone spacing increases towards the pole]. 
Gridding in this manner ensures good spatial resolution close to
disk. In the $r$ direction, we used $d{r}_{{\rm k+1}}/d{r}_{{\rm k}}=1.04$,
which enables a good resolution at smaller radii.

We chose the boundary condition at the pole (i.e., $\theta = 0^\circ$) 
as an axis-of-symmetry boundary condition. At the disk (i.e., $\theta=90^\circ $), we applied
the reflecting boundary condition. For the inner and outer radial boundaries, 
we used an outflow boundary condition (i.e., to extrapolate the flow beyond 
the boundary, we set values of variables in the ghost zones equal to 
the values in the corresponding active zones, see Stone \& Norman 1992 
for more details).

\subsection{RADIATION FIELD}
We deal with the radiation field  and radiation heating and cooling in the the same manner as described by Proga et al. (2002; see also Proga et al. 2000).

The net cooling rate is a function of the photoionization parameter, which is defined as:
\begin{equation}
\xi = \frac{4 \pi {\cal F}_{\rm X}}{n},
\end{equation}
where ${\cal F}_{\rm X}$ is the local  X-ray flux, $n ={\rho}/({m_{\rm p} \mu })$
is the number density of the gas. 
We consider fully ionized gas with $\mu=0.6$.
The  local X-ray flux is corrected for
the optical depth effects:
\begin{equation}
{\cal F}_{\rm X}={\cal F}_\ast \exp(-\tau_{\rm X}),
\end{equation}
where $\tau_{\rm X}$ is the X-ray optical depth and ${\cal F}_\ast$ is:
\begin{equation}
{\cal F}_\ast = \frac{{\rm L}_\ast}{4 \pi r^{2}}
\end{equation}
with ${\rm L}_\ast$ being the luminosity of the central source.

We estimate $\tau_{\rm X}$ between the central
source and a point in a flow from
\begin{equation}
 \tau_{\rm X}= \int_0^{r} \kappa_{\rm X} \rho(r, \theta)~dr,
\end{equation}
where $\kappa_{\rm X}$ is the absorption coefficient,
and $r$ is the distance from the central source. We assume 
%$\kappa_{\rm X}$ = 40 g $\rm{cm^{-3}}$ for $\xi \le 10^{5}$ and 
$\kappa_{\rm X}=0.4\, \rm{g^{-1}\,cm^2}$ 
%for $\xi \le 10^{5}$, 
which is numerically of value of electron scattering. 
We found that in most cases, the wind is optically thin and 
the wind column density,
\begin{equation}
N_H(\theta)= \int_{r_i}^{r_o} \frac{\rho(r, \theta)}{\mu m_{\rm p }} dr
\end{equation}
is less than $10^{23}\,\rm{cm^{-2}}$.

\section{RESULTS}

To complete the specification of the simulations, we need
to assign the properties of the disk atmosphere (the lower
boundary condition for the wind solution).
For the density, we use a simple distribution at
$\theta~=~{90}^\circ$ of the form: $\rho = {\rho_{o}} {(r/r_{{\rm IC}})}^{- \alpha}$, 
where $\alpha$ and $\rho_{o}$ are constants. We run the model for 
$\alpha$ = 0, 1, 2, and 3 and for $\rho_{o}$ = $10^{-12}$, $10^{-11}$,  and 
$10^{-10}$~g~cm$^{-3}$ (see table~1). The temperature is set to 
$10^{4} \, {\rm K}$, while the radial velocity is set to zero. In addition, 
we enforce  Keplerian rotation at $\theta=90^\circ$.
The size of the computational domain 
is defined in the following way: $r_{{\rm i}}$ = $0.05 \, R_{{\rm IC}}$ 
while $r_{{\rm o}}$ = $20 \, R_{{\rm IC}}$.
We follow M06 and N06 and adopt the following 
properties of the system: the mass of the central black hole 
$\rm{M_{BH}}$ = 7 $\MSUN$, the luminosity ${\rm L}_\ast =\, 0.03$
in units of the Eddington limit, ${\rm L}_{{\rm Edd}}$, and 
the Compton temperature,
$T_{{\rm IC}} \,=\, 1.4 \times 10^7$ K. The adopted $L_\ast$ corresponds
to the mass accretion rate $\MDOT_{{\rm a}}=4.4\times10^{17}$~g~s$^{-1}$.

\subsection{PROPERTIES OF THE FIDUCIAL RUN}

Our fiducial run (C8) has the following parameters: 
$\alpha$=2, $\rho_o = 10^{-11}\,g\,\rm{cm^{-3}}$, 
$ d{\theta}_{{\rm k+1}}/d{\theta}_{{\rm k}}=0.97$, $d{r}_{{\rm k+1}}/d{r}_{{\rm k}}=1.04$. 
This run settles to a steady state after about 3 sound-crossing times. 

We show the geometry and structure of the fiducial run in Figure~1. 
The top left panel shows a density distribution with a significant
departure from spherical symmetry.
The temperature (top right panel) is close to 0.1 $T_{{\rm IC}}$ in the outer
part of the wind. In a very narrow region around the rotational axis, 
the temperature is much less than $T_{{\rm IC}}$ at $r> 16 R_{\rm IC}$,
whereas at the small radii the temperature is comparable to
$T_{{\rm IC}}$. The X-ray irradiated gas accelerates rapidly and becomes
supersonic relatively close to the disk. As shown in the bottom left panel, 
the contour for the Mach number of one corresponding to the wind
launched at relatively large radii
is almost a straight line making an angle of about $15^\circ$ with  the disk midplane. 
Streamlines (bottom right panel) show that most of the disk wind follows 
almost perfect straight lines inclined at the angle of about $45^\circ$ with
respect to the disk midplane. The streamlines are curved
only in zones near the rotational axis and very close to the disk.  
It appears that the outer wind is self-similar
(we will return to this point at the end of this subsection).
The streamlines also show that the flow expands in the angular direction, 
reminiscent of a spherical outflow, only 
for streamlines originating at $r\simless 3~R_{\rm IC}$.

An accretion disk is geometrically extended and as such it intercepts
the central radiation over a large range of radii. However, 
the thermal expansion drives a significant wind only within 
a relatively narrow radial range.
One can show this wind property by plotting the product of the density and 
the velocity normal to the disk  as a function 
of the radius along the disk midplane (see Figure~2). Most of the outflow in 
the fiducial run comes from the narrow ring on the disk in the zone 
between $0.07 \, R_{{\rm IC}}$ and $0.2 \, R_{{\rm IC}}$ having its maximum 
at $\sim 0.1 \, R_{{\rm IC}}$. For 
radii  $> \,0.2 \, R_{{\rm IC}}$ the mass flux density
scales like $r^{-q}$, with $q=1.76$.

The wind is launched in a non-spherical way and it remains non-spherical
as it expands. To illustrate this point,
Figure~3 shows various flow properties at $r_{\rm o}$ as functions of $\theta$. 
In particular, the top panel shows the product of density and 
radial velocity (solid line) and the accumulated mass-loss rate (dotted line).
We evaluate the accumulative mass-loss rate throughout $r_{\rm o}$ using 
the following formula:
\begin{equation}
 d\dot{m}(\theta) = 4 \pi {r_{\rm o}}^2 \int_{0^\circ}^{\theta} \rho v_{\rm r} \sin \theta 
d\theta.
\end{equation} 
One finds that most of the mass flows out of the computational 
domain through the outer spherical boundary over a broad range of angles between 
$20^\circ$ and $75^\circ$. The outflow for large $\theta$ contributes 
insignificantly. 
The total mass-loss rate for the fiducial run is $\MDOT_{\rm w}= d\dot{m}(90^\circ) 
= 3.3 \times10^{18} \, \rm {g \, s^{-1}}$. 
Note that this is larger than the assumed accretion rate $\MDOT_{\rm a}$ by a factor of 7.5.
We will return to this point in Section~5.

The middle panel of Figure~3 shows that the photoionization parameter 
(solid line) decreases with increasing
$\theta$, which is another indication that the wind is not spherically 
symmetric. In particular, $\xi$ changes 
from $10^{5}$ to $10^{3}$, whereas the column density (dotted line) 
increases from below $10^{21} \, \rm {cm^{-3}} $ to $10^{23} \, \rm {cm^{-3}} $. 

The bottom panel of Figure~3 presents the radial velocity (solid line) and 
the number density (dotted line) vs. $\theta$. 
It is evident from this panel 
why the zones close to the disk and the rotational axis 
contribute insignificantly to the total mass-loss rate:
the zone near the disk is very dense but also very slow,
while the zone near the axis of rotation has extremely low density, 
so that the product of the density and radial velocity is very small. 

To specify the departure of the disk wind from a purely radial 
wind, Figure~4 shows $\xi$ (top panel) and $n$
(bottom panel) as a function of radius for various $\theta$. 
In the case of an optically thin radial wind with a constant velocity, 
$\xi$ is constant and $n \propto r^{-2}$.
In our solution, $\xi$ decreases with increasing radius especially
for small and intermediate $\theta$, whereas $n$ decreases slower 
than $r^{-2}$ and can even increase with radius for $\theta < 62^\circ$.

How does our solution compare to a $\beta$-velocity law, in terms
of approaching a constant value at large $r$? As expected, it is in agreement
for large radii. Figure~5 shows an example of the radial velocity as 
a function of radius for $\theta = 76^o$ (this corresponds to the dot-dashed 
lines in Fig. 4). The radial velocity increases up to the Compton radius and 
then remains roughly constant. Also, there is a visible dip at 
0.4 $\rm{R_{\rm IC}}$, which is a consequence of the radial line crossing 
different streamlines. The function has a nearly flat shape for radii larger 
than $1.0 \, R_{{\rm IC}}$.

We conclude that a thermal disk wind can not be approximated by
a radial outflow. Font et al. (2004) found that 
a disk wind expands quasi-spherically if a relatively steep
decline of the mass flux density with radius is assumed 
($\rho v_\theta \propto r^{-q}$, with $q\simgreat 2.5$).
As we mentioned above, in our self-consistent simulations,
$q\approx 1.76$ in the self-similar part of the wind.
It is unfortunate that the wind expansion is non-radial 
because a radial outflow is a very simple case that can be (and has been)
straightforwardly implemented into photoionization/spectra calculations.
Is it then possible to approximate the thermal wind with some other 
simple model? It is beyond the scope of this
paper to answer this question.

\subsection{COMPARISON OF THE FIDUCIAL RUN WITH OTHER RUNS}

To check the robustness of the solution presented above, we 
explored effects of the numerical resolution and the density along
the lower boundary.
We present the most significant properties of those runs in Table 1. 
The columns are organized in the following way: (1) the name of the run; 
(2) the $\alpha$ parameter used in the density profile; 
(3) $\rho_{{\rm o}}$, the normalized density at the lower boundary of 
the computational domain; 
(4) $d\theta_{{\rm i+1}}/d \theta_{\rm i}$, 
the angular resolution; (5) $\rho(r_{{\rm in}})$
the disk density at the inner radius; 
(6) $\MDOT_{\rm w}$ the total wind mass-loss rate; 
(7) the maximal angle $\theta_{{\rm max}}$ for which the integration of 
the total wind mass-loss is computed [$\MDOT_{\rm w}= d\dot{m}(\theta_{{\rm max}})$];
(8) the maximal radial velocity on the outer shell; 
(9) the total time of the run.

\begin{table*}
\scriptsize
\begin{center}
\caption{ Summary of the results:}
\begin{tabular}{c c c c c c c c c} \\ \hline
\hline
No. of run  & $\alpha$ & $\rho_{o}$  & $d\theta_{i+1}/d \theta_i$ &$\rho\,(r_{in})$ &     $\rm{\MDOT}_{wind}$     & $\theta_{max}$ &  $v_{r}$ & ${t_f}$\\
    & & ($\rm {g \,cm^{-3}}$)  & &($\rm {g \,cm^{-3}}$)    & ($10^{18}\, \rm { g \,s^{-1}}$) & ($^\circ$)  & ($\rm {\,km\, s^{-1}}$) & ($2.2 \times 10^5 \, \rm { s}\,\,^{*}$) 
\\\hline
 A   & 0 &$10^{-10} $ &0.97&$10^{-9 }$ & 0.33  & 59  &587 & 10 \\

 &&&&&&&& \\

 B   & 1 &$10^{-10} $ &0.97&$10^{-9 }$ & 0.33  & 67  &599& 10 \\

 &&&&&&&& \\

 C1  & 2 &$10^{-12} $ &1.00&$10^{-10}$ & 6.6  & 90   &746& 10 \\

 C2  & 2 &$10^{-12} $ &0.99&$10^{-10}$ & 6.8  & 90   &708& 10 \\
 
 C3  & 2 &$10^{-12} $ &0.98&$10^{-10}$ & 6.9  & 90   &675& 10 \\
 
 C4  & 2 &$10^{-12} $ &0.97&$10^{-10}$ & 7.0  & 90   &642& 10 \\
 
 C5  & 2 &$10^{-12} $ &0.96&$10^{-10}$ & 7.1  & 90   &619& 10 \\
 
 C6  & 2 &$10^{-12} $ &0.95&$10^{-10}$ & 7.1  & 90   &604& 30 \\
 
 C7  & 2 &$10^{-11} $ &0.99&$10^{-9 }$ & 0.32 & 90   &703& 10 \\

 C8  & 2 &$10^{-11} $ &0.97&$10^{-9 }$ & 0.33  & 90  &627& 10 \\

 C9  & 2 &$10^{-11} $ &0.95&$10^{-9 }$ & 0.33  & 90  &587& 10 \\

 C10 & 2 &$10^{-10} $ &0.99&$10^{-8 }$ & 0.33  & 63  &692& 30 \\

 C11 & 2 &$10^{-10} $ &0.97&$10^{-8 }$ & 0.33  & 62  &603& 30 \\

 C12 & 2 &$10^{-10} $ &0.95&$10^{-8 }$ & 0.33  & 63  &553& 10 \\

 &&&&&&&& \\

 D1  & 3 &$10^{-11} $ &0.97&$10^{-8 }$ & 0.11  & 90  &670& 10 \\

 D2  & 3 &$10^{-10} $ &0.97&$10^{-7 }$ & 0.32  & 90  &621& 30 \\

\\\hline
\multicolumn{9}{l}{* $2.2 \times 10^5 \, \rm { s}$ is the sound crossing time = $(r_o/c_s$), calculated for T=$T_{\rm IC}$. }
\end{tabular}

\end{center}
\normalsize
\end{table*}

As mentioned in the Section 3, the density along the lower boundary 
($\theta=90^\circ$), is specified by two parameters, $\alpha$ and $\rho_{\rm o}$. 
We have run several models for various $\alpha$ and $\rho_{\rm o}$ to check
if our solution depends on the density along the disk.
As for radiation driven disk winds (see e.g., Proga et al. 1998),
we find that as long as the density along the disk is high enough,
the gross properties of the thermal disk wind do not depend on the assumed
disk density.

For a relatively small density along the disk, the X-rays heat the gas 
to the Compton temperature even at $\theta=90^\circ$. This means that
the computational domain does not capture a cold disk and its part that is
in hydrostatic equilibrium. Consequently, the simulations do not 
follow the transition between a cold and a hot flow.
Therefore, the mass flux density from the lower boundary is set not
by the physics of a cold disk being heated by X-rays  but by 
a choice of the density.
Comparing $\MDOT_{\rm w}$ for runs C4, C8, C11, and D2,
we find that the mass-loss rate depends on the density if the density
is too low. Only when the density along the disk is high enough
that the central X-rays do not heat the gas along the lower boundary,
the simulations capture the transition from the cold to hot phase.
The mass-loss rate then becomes independent of a particular choice
of the density in the cold part of the disk.

However, we also find that for very high densities along the disk
there is a technical problem with a proper measurement of $\MDOT_{\rm w}$.
Our goal is to compute a wind from a geometrically thin disk.
But if we choose too a high density along the lower boundary,
the disk will flare and its thickness at large radii becomes
substantial. During the simulations, the dense disk 
remains cold and nearly in hydrostatic equilibrium but
it will subsonically fluctuate. Therefore, when computing the {\it wind} 
mass-loss rate, one needs to keep in mind these fluctuations of the dense disk 
and exclude the region very close to the disk midplane. Otherwise
the subsonic oscillations of the very dense disk material would be counted as
the disk outflow and this will yield an erroneous estimate of the wind mass
loss rate. 
In practice, when calculating the total outflow rate for some 
runs, we stop the integration in eq. 10 
at $\theta_{{\rm max}}\,<\,90^o$, 
where the transition from the wind to the disk occurs. This transition
is not difficult to identify as it occurs where the density
sharply increases  with $\theta$ (see the bottom panel in Fig. 3). 
The seventh column in Table~1 lists the maximum angles we used. 

For the fiducial run as well as some other runs, the region near the disk midplane 
does not affect estimations of $\MDOT_{\rm w}$. This is one of the reasons 
we chose the run C8 with
$\rho_{\rm o}$ = $10^{-11}\, \rm {g \,cm^{-3}}$ as the most suitable run
in terms of having not too high a density, as only a small portion 
of the dense disk  will enter the domain. On 
the other hand, the density cannot be so small as to make
the transition from the disk 
to the wind outside the computational domain.
All runs with the density comparable to or larger than that in the fiducial run
have a total mass-loss rate of about 3.3 $\times \,10^{17}\, \rm { g \,s^{-1}}$. 
Other wind properties are also similar (see Table~1).

As for the resolution study, we find that
changes in the $ d{\theta}_{{\rm k+1}}/d{\theta}_{{\rm k}}$  factor 
produced small differences of the final steady state and 
such effects just slightly changed calculated physical quantities and 
properties of the solution.

\section{COMPARISON WITH OBSERVATIONS}

We start our comparison with the observations by considering the wind
properties inferred from the observation and then we present
an example of the synthetic spectrum directly compared
with the observed X-ray spectrum.
As we described in Section~1, the number density of the wind
is inferred to be quite high, i.e., $5\times 10^{15}$~cm$^{-3}$.

Figure~6 shows the scatter plot of the photoionization parameter vs. 
$v_{{\rm r}}$ based on run C8. 
Dots correspond to
$n < 10^{12}~{\rm {cm}}^{-3}$, whereas diamonds correspond
$n \geq 10^{12}~{\rm {cm}}^{-3}$.
Note that there are {\it no diamonds} corresponding to
$v_{{\rm r}} \geq \rm{100 \,km\, s^{-1}}$. This means that our model 
fails to  predict the wind density and velocity as inferred from
the observations. We stress that in the computational domain of
our simulation  there is gas with densities as high as $n \approx 5\times10^{15}$~cm$^{-3}$, but
this high density gas is very slow (e.g., the bottom panel of Fig.~3).
This simple comparison shows that our results
support M06 and Kallman et al.'s (2009) conclusion:
a thermal wind can not account for the X-ray observations.

As another test of our model, we computed synthetic spectra
using the photoionization code XSTAR (Kallman \& Bautista 2001; 
Bautista \& Kallman 2001). Our method of making this 
calculation is very similar to that used and described  
in Dorodnitsyn et al. (2008). We have computed spectra of the fiducial model for several
inclination angles. We have also
calculated the emissivity and opacity at each point in 
the hydrodynamic flow at the end of the simulations, and then integrated 
the formal solution of the transfer equation along radial rays to get 
the spectrum.
In doing this we ignore emission.  Since emission  will tend to fill in 
absorption lines, thus reducing their strength, these models represent an
overestimate to the amount of absorption coming from such a flow.
For most inclinations, the spectra show
no absorption features or only very weak ones. Figure~7 compares
the count spectrum observed by the HETG
during an X-ray bright phase of the GRO J1655--40 2005 outburst (see
M06 for more details)
with an example of our synthetic spectrum.

For this figure, we have chosen an almost $90^\circ$ inclination 
because it yields the best fit to the data and 
the right ionization for the most lines. 
Nonetheless, even for this very high inclination,
the range of ionization is too narrow, and 
the ionization is probably lower than that observed. 
It is possible to obtain higher ionization for
lower inclinations, but then the column density is lower, 
making the fit is even worse. To see if we can improve the fit
for a given density structure,
we have to put in the 600 km ${\rm{s}^{-1}}$ blueshift by hand while computing 
this spectrum (note that our wind model does not predict such a high velocity 
of the dense wind at this inclination angle). However,
neither this nor any other simple model fits the Fe K$\alpha$ line 
profile showing the high velocity ($\sim$1600 km ${\rm{s}^{-1}}$) 
component. We found also that the curve-of-growth of the lines does not fit: 
there are  no saturated lines in the model. Still, 
close to half of the observed lines are predicted in the 
model, including the Fe XXII lines.

In addition to an inadequate mass loss rate, the model
shows a geometric structure different from that required
by spectra of GRO J1655--40.  From the lack of emission
lines, M06 and N06 concluded that the
\begin{comment}
emission (global) covering factor
\end{comment}
global covering factor
of the wind is small, while Figure 3 shows that the density
is reasonably constant over a broad range of angles.

\section{SUMMARY AND THE DISCUSSION}

In this paper, we presented axisymmetric, time-dependent hydrodynamical 
numerical simulations of thermally driven disk winds. 
To apply our results to GRO J1655--40, we adopted the observed mass of 
its black hole, and the
properties of its X-ray radiation. We performed the simulations
using the same code and in a similar fashion to PK02 who
computed wind for low-mass X-ray binaries. The main difference
is that we turned off the radiation driving here because
as found by PK02 it is negligible for the stellar black hole
accretors. Turning off the radiation driving makes the simulations run
much faster.
The main goal of our simulations is to check whether, in the absence of 
magnetic fields and within the given simplifications and assumptions, 
the thermally driven wind can account for the observation. 

To compare the results with the observations, we also computed transmitted 
X-ray spectra using the photoionization code XSTAR and our wind solution.
Our main findings are:\\
1) The density of the fast-moving part of the wind is more than one order 
of magnitude lower than that inferred from the observations. 
Consequently,  contrary to the claim made by N06, the thermal model fails 
to predict synthetic spectra with 
line absorption as strong and as blueshifted as those observed. 
Overall our results support the conclusion reached by Miller et al. (2008)
and Kallman et al. (2009), that GRO J1655--40, and 
likely other  X-ray transient sources, have thermal winds insufficient 
to explain the observed spectra; \\
2) Despite the thermal wind being weak and 
Compton thin, the ratio between  the mass-loss rate and the mass accretion 
rate is about seven;\\ 
3) We discovered that the outer wind is self-similar.

One should ask if our simulations are conclusive. Namely,
can we rule out thermal driving as a mechanism responsible for 
the observed wind even though we
did not consider some effects such as thermal conduction?
Conduction could increase the degree of disk heating 
and could in principle increase the gas density above the disk photosphere.
However, it is unclear if this increase
would lead to an increase in the density at the wind base
and consequently in an increase of
the density of the fast wind,
which is what is needed to account for the observations.
The density of the wind base also could be sensitive to 
the details of radiative transfer and spatial resolution.
Hydrostatic models of the ionization
layer on the surface of the disk by Jimenez-Garate et al.
(2002, 2005) showed that there
is much more opacity than
the electron scattering opacity
in the very thin layer that produces UV and most of the X-ray
emission on top of the optically thick disk. Therefore,
our pure electron scattering assumption might underestimate
the opacity. However, if this were true then that would mean
that thermal driving is even less efficient because
the irradiation would penetrate the disk even less.
We expect that the mass loss rate is set at the sonic
point, which would occur in the million Kelvin temperature
plateau seen in the hydrostatic models.  In that region,
the models indicate that photoabsorption opacity due to
Fe XXV and Fe XXVI is comparable to the electron scattering
opacity.
The radiative heating also can be affected by disk flaring
or the shape of the disk, in general. For example, at a given radius 
a strongly flared disk might intercept more central radiation 
than a flat disk. Consequently, a resulting outflow might be more
collimated and denser. We intend to extend our model to explore these effects 
in near future. However, before that we plan to check if the inclusion of 
magnetic driving will produce a dense fast wind.

It is important to remember that the outburst spectra of
GRO J1655--40 and GRS 1915+05 are exceptional.  Spectra of
these objects in their low states, as well as spectra of
other black hole binaries, show absorption lines of only Fe
 XXV and Fe XXVI.  That suggests lower densities and
mass loss rates, and it raises the question of whether thermally
driven winds might account for those observations.  The
microquasar H1743$-$322 (Miller et al. 2006) is a good example,
with Fe XXV and Fe XXVI equivalent widths of up to 4.3 m\AA\/
and 6.8 m\AA , respectively.  The densities are below $10^{13}~\rm cm^{-3}$
and the ionization parameters are above $10^5$ in 4 separate observations.
We have examined our models at sightlines farther from the disk,
where the velocities match those reported by Miller et al. (2006).
The Fe XXV and Fe XXVI column densities are too small to account
for the observed equivalent widths by a factor of a few, and we
conclude that thermally driven winds cannot account for the
typical low state winds. 

In the models presented here, we found that 
$\MDOT_{\rm w}/\MDOT_{\rm a} \approx 7$. This is a fairly large
ratio, which could mean that the thermally driven wind can significantly
change the mass flow in the disk. Therefore, we decided to check if our result 
is consistent with the results of others who modeled thermally driven
winds. Of particular interest is work by Woods et al. (1996), who
studied thermally driven disk winds in great detail and performed
many axisymmetric wind simulations for various luminosities.
They summarized their results in 
an analytic formula which fitted the mass flux density 
distribution obtained from simulations  for  various $L_\ast$
(their eq. 5.2).

To find $\MDOT_{\rm w}\, / \,\MDOT_{{\rm a}}$  predicted by Woods et al.'s 
simulations, we integrated their formula over the disk surface to obtain 
$\MDOT_{{\rm wind}}$ and then expressed $L_\ast$ 
through the following equation of the accretion luminosity:
\begin{equation}
L_\ast = \frac{\rm{G\,M_{BH}\,\dot{M}_{a}}}{2\,r_{\rm s}},
\end{equation}
where $r_{\rm S}$ is the Schwarzschild radius, which yields a relation 
between $\dot{M}_w$ and $\dot{M}_a$. 
We found that Woods et al. simulations predict $\MDOT_{{\rm w}}\, / \,\MDOT_{{\rm a}}$  
in a range from 2.0 to 6.0 depending on the luminosity. The $L_\ast$ 
dependence is weak for $L_\ast / L_{{\rm Edd}}<0.1$. Thus 
$\MDOT_{\rm w}\, / \,\MDOT_{\rm a}$ 
is higher than, but comparable, to Woods et al.'s results for $L_\ast$ on 
the lower side of their luminosity range. Nevertheless, both
set of simulations predict $\MDOT_{\rm w}\, / \,\MDOT_{\rm a}>1$. 

Our conclusion from this analysis is that $\MDOT_{\rm w}$ is greater than the rate 
at which the central engine is fueled. This in turn can make the disk variable. 
One could even expect this relatively strong wind to cause a recurrent 
disappearance of the inner disk even if the disk is fed at a constant rate
at large radii.
Neilsen \& Lee (2009) suggested such a process while interpreting 
jet/wind/radiation variability in the microquasar GRS 1915+105. 
It is beyond the scope of this work to model the effect of the wind 
on the disk. However, this problem was studied by
Shields et al. (1986), who showed that due to viscous processes 
no oscillations appear for $\MDOT_{\rm w}\, / \,\MDOT_{\rm a}\, <\,15$!
We conclude that the thermal wind is to weak to cause the oscillation. But 
the thermal wind is also too weak to account for the observed wind.
Therefore, it is possible that in GRO J1655--40, and other sources, 
e.g., GRS 1915+105, the wind responsible for the observed X absorption
will be so strong that $\MDOT_{\rm w}\, / \,\MDOT_{\rm a}\, >\,15$
and as such cause disk oscillations and contribute to the observed
disk variability.

We thank Tim Waters for his comments on the manuscript.
We acknowledge support provided by the Chandra awards TM8-9004X and
TM0-11010X issued by 
the Chandra X-Ray Observatory Center, which is operated by the Smithsonian 
Astrophysical Observatory for and on behalf of NASA under contract NAS 8-39073.

\clearpage

\section*{ REFERENCES}
 \everypar=
   {\hangafter=1 \hangindent=.5in}

{
 
  Balbus  A., \&  Hawley J.F. 1998, Rev. Mod. Phys. 70, 1–53

  Bautista, M., \& Kallman, T. 2001, ApJS, 134, 139  

  Begelman, M.C., McKee, C.F., \& Shields, G.A. 1983, ApJ, 271, 70

  Blandford, R.D., \& Payne, D.G. 1982, MNRAS, 199, 883

  Dorodnitsyn, A., Kallman, T., \& Proga, D. 2008, ApJ, 687, 97

  Font, A. S.,  McCarthy, I. G., Johnstone, D., \& Ballantyne, D. R. 2004, ApJ, 670, 890

  Jimenez-Garate, M.A., Raymond, J. C., \& Liedahl, D.A. 2002, ApJ, 581, 1297

  Jimenez-Garate, M.A., Raymond, J.C., Liedahl, D.A., Hailey, C.J. 2005, ApJ, 
 625, 931

  Kallman, T. R., \& Bautista, M. 2001, ApJS, 133, 221

  Kallman, T. R., Bautista, M.A., Goriely, S.,  Mendoza C., Miller, J.M., 
Palmeri, P., Quinet, P., \& Raymond, J. 2009, ApJ, 701, 86

  Kubota, A., Dotani, T., Cottam, J., Kotani, T., Done, C., Ueda, Y., Fabian, A. C., Yasuda, T., Takahashi, H., Fukazawa, Y., Yamaoka, K.,; Makishima, K., Yamada, S., Kohmura, T., \& Angelini, L. 2007, PASJ, 59S, 185 

  Miller, J. M., Raymond, J., Fabian, A. C., Homan, J., Nowak, M. A., Wijnands, R., van der Klis, M., Belloni, T., Tomsick, J. A., Smith, D. M., Charles, 
P. A., \& Lewin, W. H. G. 2004, ApJ, 601, 450

  Miller, J. M., Raymond, J., Fabian, A. C., Steeghs, D., Homan, J., Reynolds, C.S., van der Klis, M., \& Wijnands, R. 2006, Nature, 441, 953 (M06)

  Miller, J. M., Raymond, J., Homan, J., Fabian, A. C., Steeghs, D., Wijnands, R., Rupen, M., Charles, P., van der Klis, M., \& Lewin, W. H. G. 2006, ApJ, 646, 394

  Miller, J.M., Raymond, J., Reynolds, C.S., Fabian, A.C., Kallman, T.R. ,
\& Homan, J. 2008, ApJ, 680, 1359

  Netzer, H. 2006, ApJ, 652, L117 (N06)

  Neilsen, J., \&  Lee, J.C. 2009, Nature, 458, 481

  Orosz, J.A., \& Bailyn, C.D. 1997, ApJ, 477, 876

  Proga, D. 2006, Nature, 441, 938

  Proga, D., \& Kallman, T. 2002, ApJ, 565, 455 (PK02)

  Proga, D., Stone J.M., \& Drew J.E. 1998, MNRAS, 295, 595

  Proga, D. , Stone, J., \& Kallman, T. 2000, ApJ, 543, 686

  Shields, G.A., McKee, C.F., Lin, D.N.C., \& Begelman M.C. 1986, ApJ, 306, 90

  Shulz, N.S., \& Brandt, W.N. 2002, ApJ, 572, 971

  Stone, J.M., \& Norman, M.L. 1992, ApJ Suppl., 80, 753

  Woods, D.T., Klein, R.I., Castor, J.I., \& McKee, C.F. 1996, ApJ, 461, 767

}

\eject
%\begin{comment}

\begin{figure}
\begin{picture}(330,545)
\put(-200,-220){\includegraphics{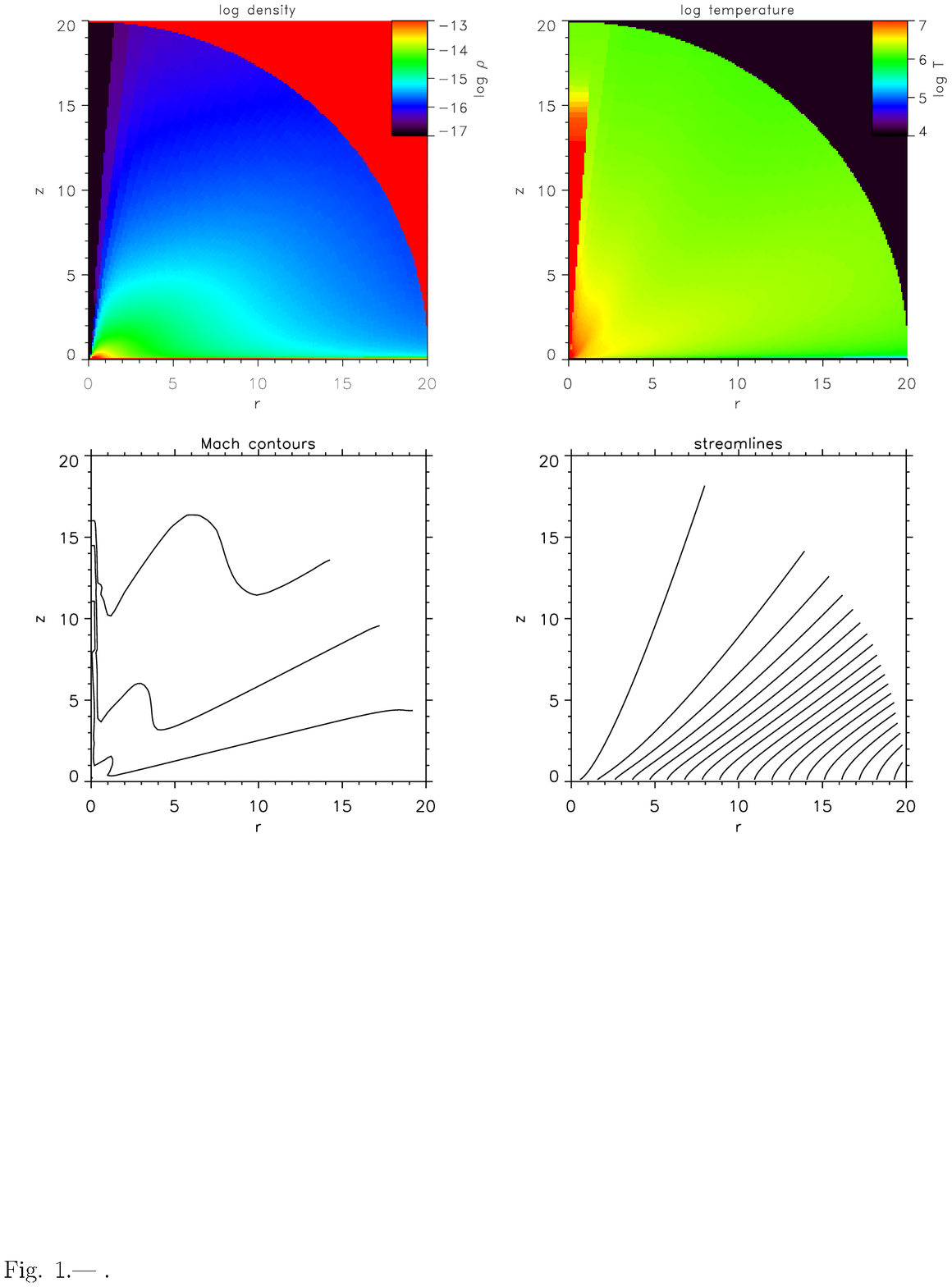}}
\end{picture}
\caption{ The fiducial run C8. \textit{Top left panel: } Color density map.  
\textit{Top right panel:} Color temperature map. 
\textit{Bottom left panel: } The Mach number M contours (the poloidal 
component only). Contours are for M = 1, 2, and 3 (bottom to top).
\textit{Bottom right panel: } The flow streamlines. Note
the self-similarity of the streamlines especially those
arising from the outer disk.
In all panels the rotation axis of the disk is along 
the left-hand vertical frame, while the midplane of the disk is along 
the lower horizontal frame. Lengths are expressed in units of the Compton
radius, $R_{\rm IC}$.
}
\end{figure}
%\end{comment}

\begin{figure}
\begin{picture}(180,180)
\put(330,-145){\includegraphics{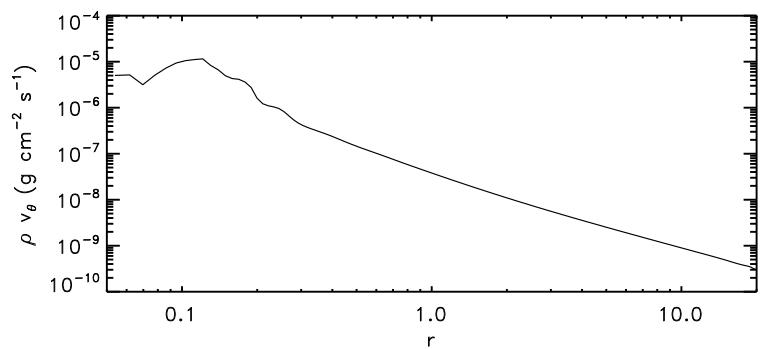}}
\end{picture}
\caption{ The mass flux density, $\rho  v_{\theta}$, as the function of radius 
along the disk midplane, $\theta  = 90^\circ$ for the fiducial run.
}
\end{figure}

\begin{figure}
\begin{picture}(330,439)
\put(330,150){\includegraphics{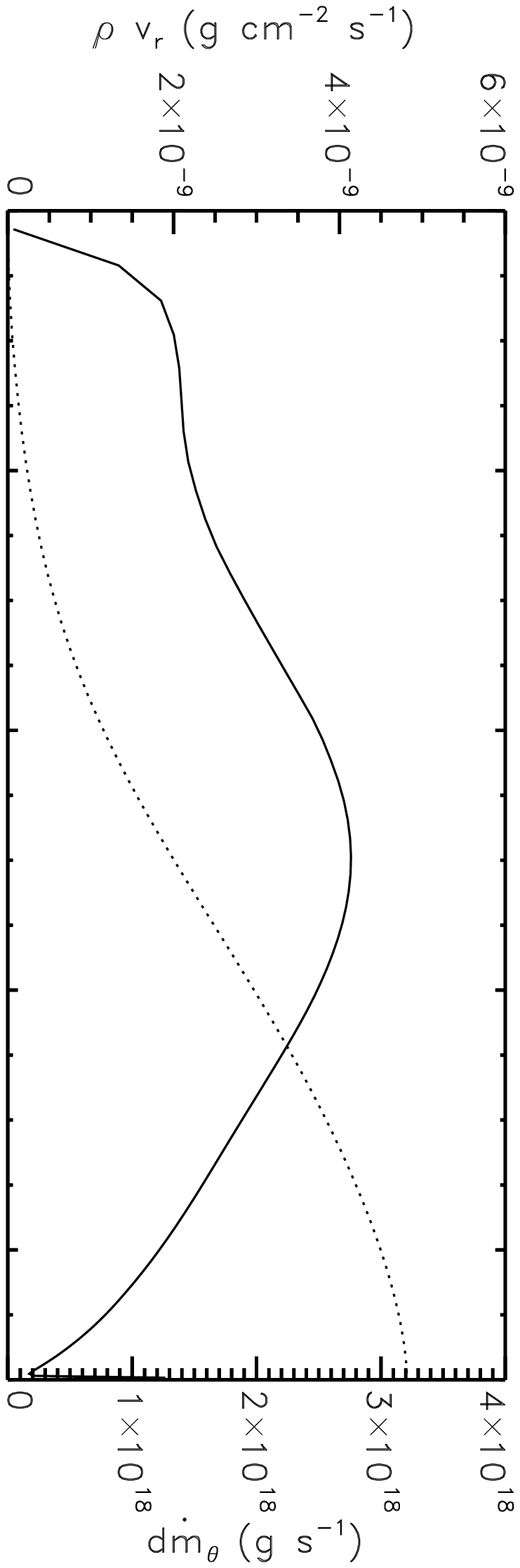}}
\put(330,30){\includegraphics{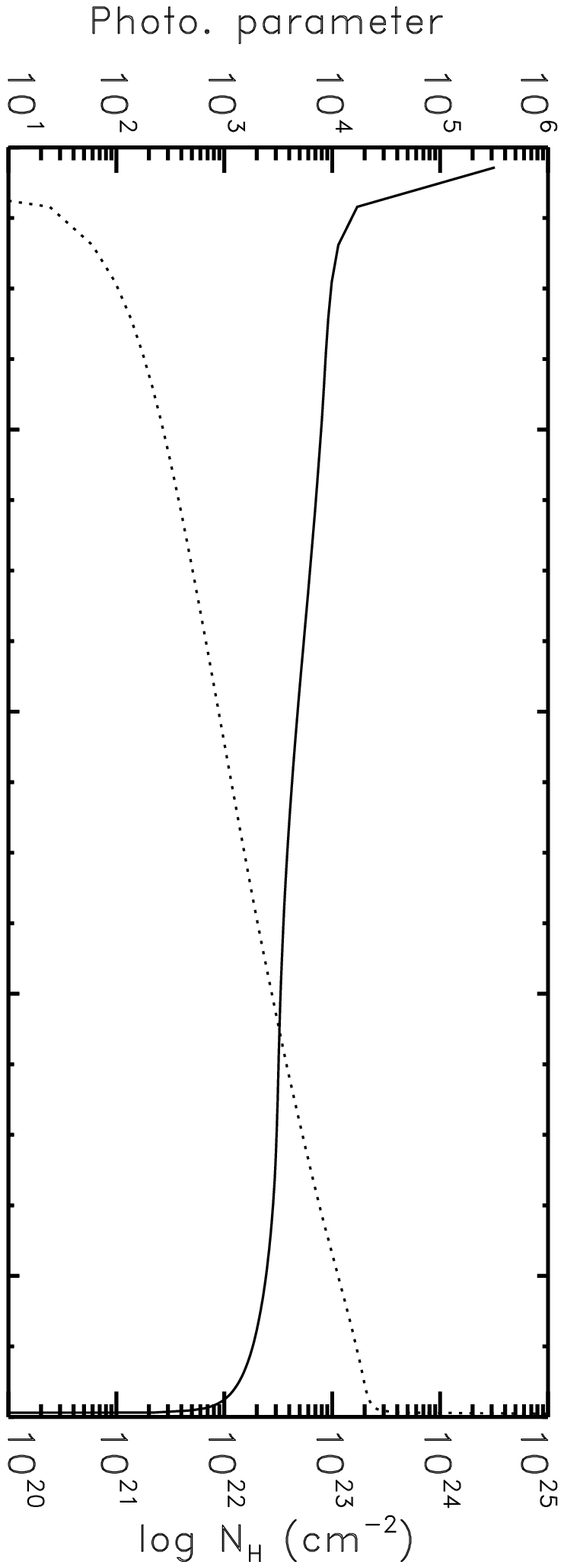}}
\put(330,-90){\includegraphics{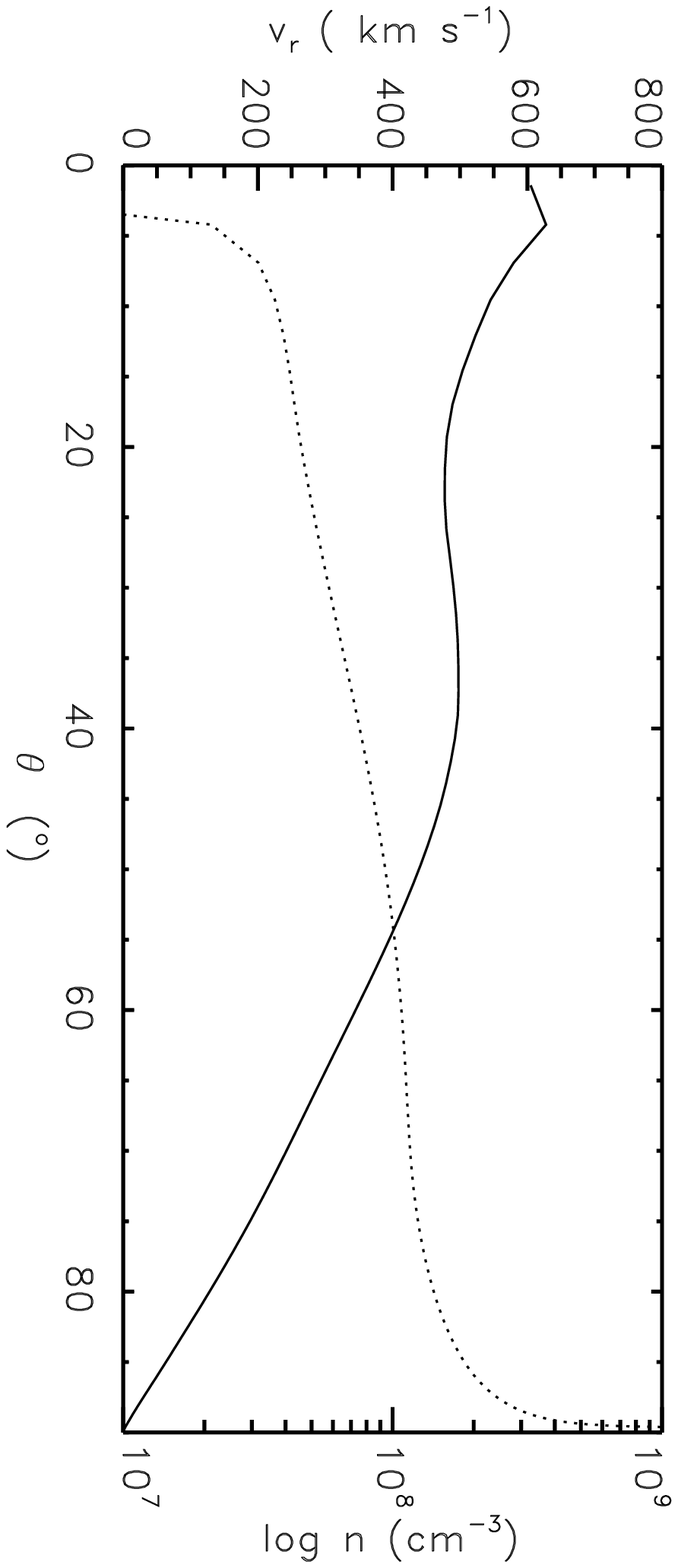}}
\end{picture}
\caption{  Quantities at the outer boundary, $r_{\rm o}$ of the fiducial model.  The ordinate on the left hand side refers to the solid line, while 
the ordinate on the right hand side refers to the dotted line. 
\textit{Top panel:} Mass flux density and the accumulative mass-loss rate. 
\textit{Middle panel: }Photoionization parameter and the column density. 
\textit{Bottom panel: } radial velocity and number density.
}
\end{figure}

%\begin{comment}

\begin{figure}
\begin{picture}(180,240)
\put(330,-100){\includegraphics{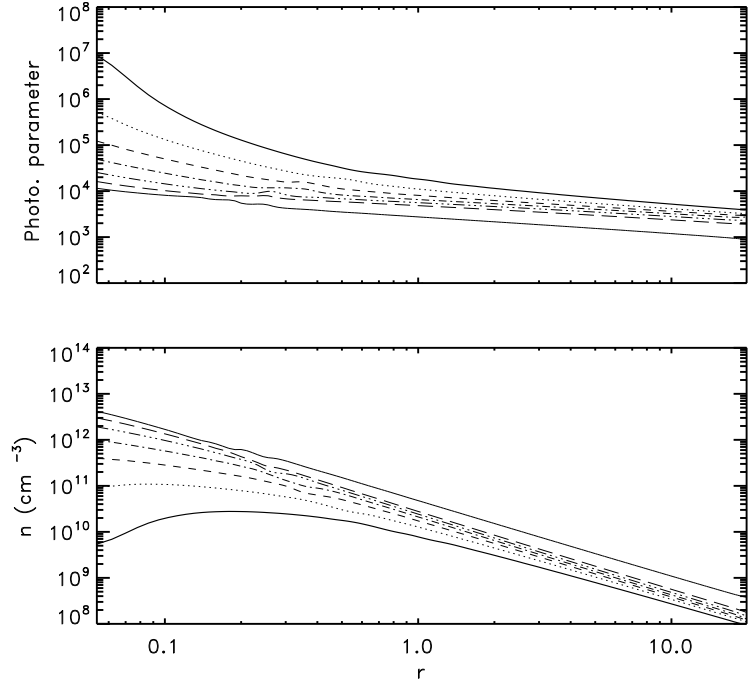}}
\end{picture}
\caption{  
The radial profiles of the photoionization parameter (top panel) and 
the number density (bottom panel) for seven polar angles along the radius
for the fiducial run.
The angles are:
 $\theta=48.3^\circ$ (thick solid),
 $\theta=60.5^\circ$ (dotted),
 $\theta=69.4^\circ$ (dashed),
 $\theta=76.0^\circ$ (dot-dashed),
 $\theta=80.9^\circ$ (triple dot-dashed),
 $\theta=84.5^\circ$ (long dashed) and
 $\theta=89.1^\circ$ (thin solid).
Note that a simple spherically expanding outflow poorly approximates 
the disk wind solution (see the main text for more details).
}
\end{figure}

\newpage

\begin{figure}
\begin{picture}(180,90)
\put(330,-150){\includegraphics{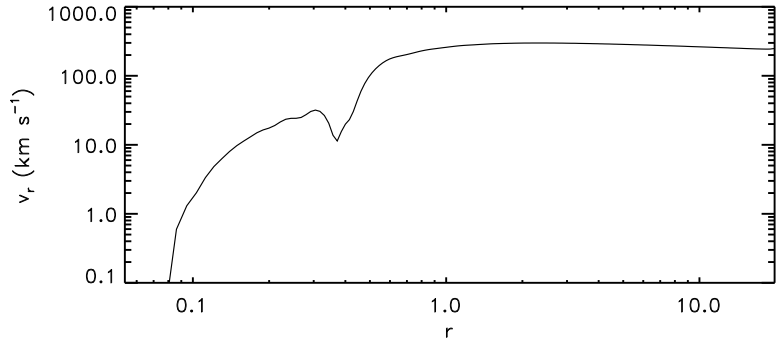}}
\end{picture}
\caption{Radial velocity as the function of radius at $\theta  = 76^\circ$
for the fiducial run. The non-monotonic radial profile 
demonstrates that the disk wind is not radial (see also Fig. 4).
}
\end{figure}

\clearpage 
\begin{figure}
\begin{picture}(330,150)
\put(330,-150){\includegraphics{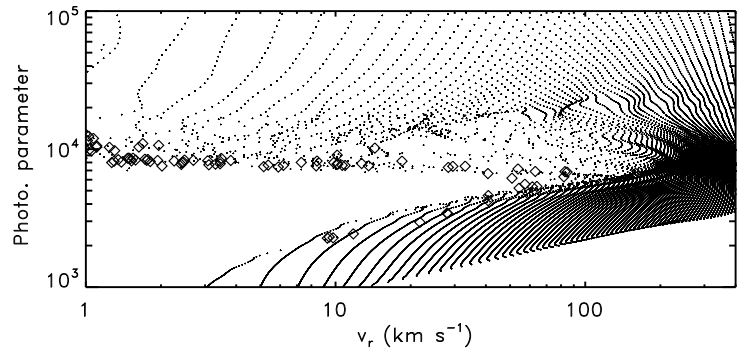}}
\end{picture}
\caption{  Scatter plot of photoionization parameter vs. $v_{r}$
for the fiducial run. 
Dots correspond to
$n < 10^{12}~{\rm cm}^{-3}$ whereas diamonds correspond
$n \geq 10^{12}~{\rm cm}^{-3}$.
The lack of points corresponding to $v_r> \rm 100~{\rm{km}~{\rm{ s}}^-1}$ 
means that the thermal
wind can not account for a high density fast outflow as observed
in GRO~J1655--40 during an X-ray-bright state.
}
\end{figure}
%\end{comment}

\clearpage

\begin{figure}
\begin{picture}(330,440)
\put(4,-50){\includegraphics{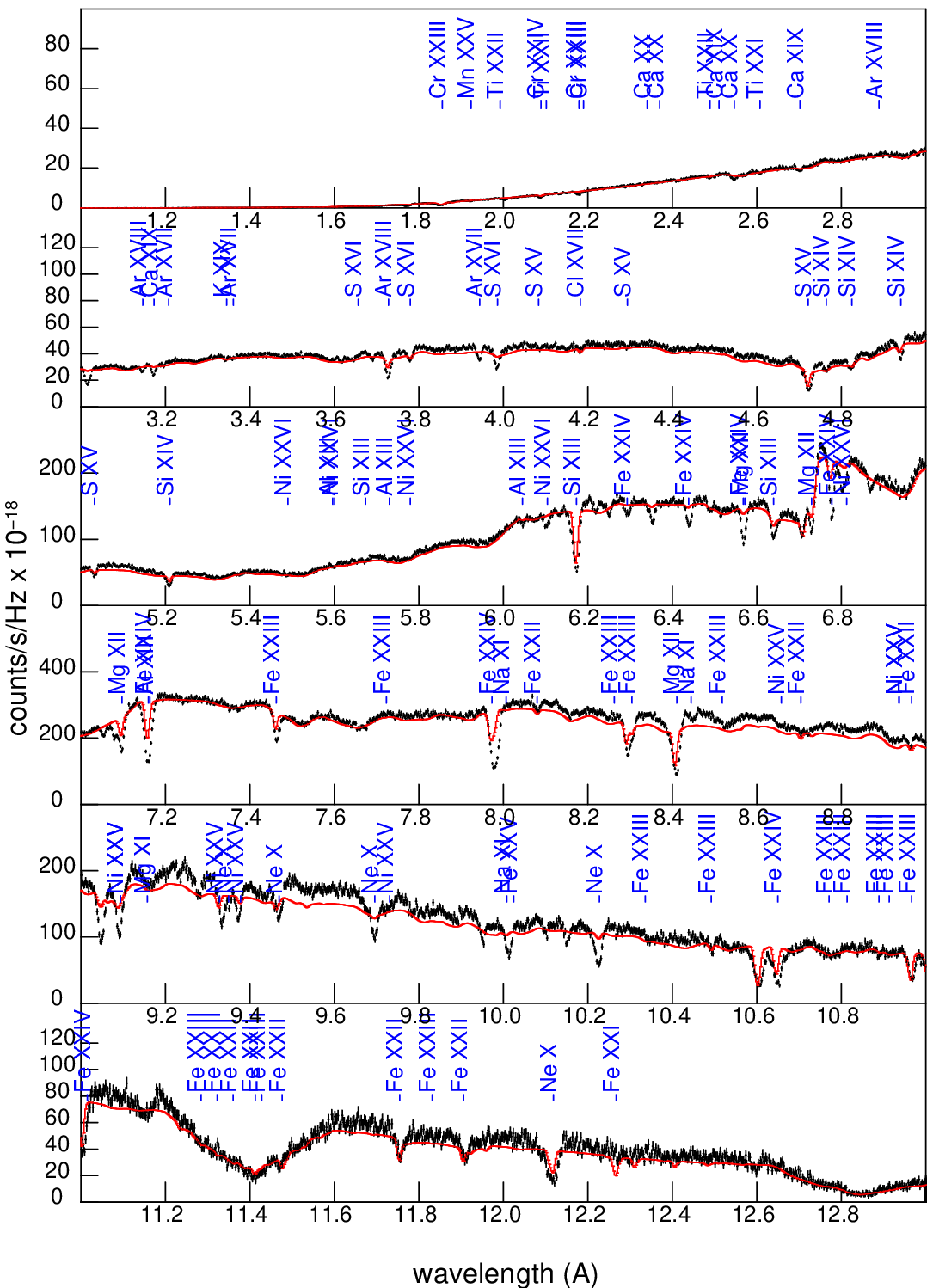}}
\end{picture}
\caption{ Comparison of the X-ray spectrum of GRO J1655--40
with the model spectrum computed based on our thermal disk wind
(see Section 4). The points with the error bars are the data and the 
model is the red curve. The wavelength in Angstroms.}
\end{figure}
\clearpage

%\plotone{picture}

%\begin{picture}(330,345)
%\put(-200,-220){\special{psfile=figure1.ps angle =00
%hoffset=130 voffset=-15 hscale=99 vscale=99}}

\end{document}